\documentclass[aps,pra,twocolumn,floatfix,showpacs]{revtex4}

\usepackage{graphicx}
\usepackage{epstopdf}
\usepackage{dcolumn}
\usepackage{bm}
\usepackage{amsmath}
\usepackage{amsxtra}
\usepackage{amssymb}
\usepackage{amsthm}

\begin{document}

\preprint{AUR1028}

\title{Metrology of time-domain soft X-ray attosecond pulses and re-evaluation of pulse durations of three recent experiments }

\author{Xi Zhao$^{1}$, Su-Ju Wang$^{1}$, Wei-Wei Yu$^{1,2}$, Hui Wei$^{1}$,  and C.~D.~Lin$^{1}$}
\affiliation{$^{1}$ Department of Physics, Kansas State University,
Manhattan, KS 66506, USA\\
$^{2}$ School of Physics and Electronic Technology, Liaoning Normal University, Dalian 116029, People's
Republic of China}

\begin{abstract}
Attosecond pulses in the soft-X-ray (SXR) to water-window energy region offer the tools for creating and studying target specific localized inner-shell electrons or holes in materials,   enable monitoring or controlling charge and energy flows in a dynamic system on attosecond timescales. Recently, a number of laboratories have reported generation of continuum harmonics in the hundred-electron-volt to kilovolt region with few-cycle long-wavelength mid-infrared lasers. These harmonics have the bandwidth to support pulses with duration of few- to few-ten attoseconds. But harmonics generated in a gas medium have attochirps that cannot be fully compensated by materials over a broad spectral range; thus, realistically what are the typical shortest attosecond pulses that one can generate? To answer this question, it is essential that the temporal attosecond pulses be accurately characterized. By re-analyzing the soft X-ray harmonics reported in three recent experiments \cite{chang_natcom2017,Thomas_OE2017,Bieger_2017PRX} using a newly developed broadband phase retrieval algorithm, we show that their generated attosecond pulses are all longer than about 60 as. Since broadband pulses tend to have high-order chirps away from the spectral center of the pulse, the   algorithm has to be able to retrieve accurately the phase over the whole bandwidth. Our re-evaluated pulse durations are found to be longer than those previously reported. We also introduce the autocorrelation (AC) of the streaking spectrogram. By comparing the ACs from the experiments and from the retrieved SXR pulses, the accuracy of the retrieved results can be directly visualized to ensure that correct phases have been obtained. Our retrieval method is fast and accurate, and it shall provide a powerful tool for the metrology of few-ten-attosecond pulses in the future.

\end{abstract}


\maketitle

\section{Main}

Since the first report of isolated attosecond pulses (IAP) in 2001 \cite{Krausz_2001nature}, the majority of IAPs are generated only in EUV (or XUV) region, with photon energy below about 120~eV \cite{Krausz_2002nature,Linbook,Krausz_2013nature}. To study temporal electron dynamics, bond breaking, and energy flow in a bio-chemical reaction, extension of IAPs to soft-X-ray region (SXR) is highly desirable since SXR would excite inner-shell electrons to create a localized initial hole, which would fingerprint the flow of charges (electrons or holes) and energy, giving access to chemical processes at the most fundamental level with best spatial and temporal resolutions.

Within the last decade, the development of high-energy long-wavelength driving lasers together with pulse compression to few cycles has allowed experimentalists  to generate broadband continuum high-energy photons up to 1.6 keV~\cite{Dimauro_2008nat_phys, Stagira_2007OL, Kapteyn_2012science, Midorikawa_2008PRL,Itatani_2014nat_phys, Biegert_2014OL, Schmidt_2010APL, Kartner_2016JPB, Biegert_2015Nat_comm,Ishii_2014Nat_comm, Keathley_2016JPB, Biegert_2015Nat_comm2, Li_2016APL}. These pulses have spectral bandwidth to support transform-limited pulses of a few attoseconds or even zeptoseconds. However, to claim such short durations would require accurate characterization of the spectral phase over the whole spectral bandwidth.

Recently, using two-cycle 1.80 to 1.85 $\mu$m lasers, three groups have reported broadband SXR harmonics~\cite{chang_natcom2017,Thomas_OE2017,Bieger_2017PRX}. In all three experiments, streaking spectra were also measured. Two groups reported pulse durations of 53$\pm 6$ as~\cite{chang_natcom2017} and 43$\pm1$ as~\cite{Thomas_OE2017}, with spectral bandwidth of 100-300 eV and  65-150 eV, respectively. Both are the shortest attosecond pulses in the world at the time. In another experiment, an upper limit of 322 as (bandwidth  150-350 eV) was reported~\cite{Bieger_2017PRX}. These pulse durations were all obtained from the spectral phases retrieved from the streaking spectra. Do we have an accurate algorithm to retrieve the phase for such broadband pulses? Is there a way to check whether the retrieved phase is indeed correct? In this contribution, with a new theoretical approach and algorithm, we demonstrate that our new method can accurately retrieve the spectral phase of a broadband pulse, and more importantly, we can check if phases have been accurately retrieved. We then use our new method to evaluate the SXR attosecond
pulses reported in ~\cite{chang_natcom2017,Thomas_OE2017,Bieger_2017PRX}.

A SXR pulse in the time domain is easily obtained via Fourier transform if the electric field in the energy domain $E_{\text{SXR}}(\Omega)=U(\Omega)e^{i\Phi(\Omega)}$ is available. The spectral amplitude $U(\Omega)$, where $\Omega$ is the photon energy, can be obtained from photoelectron spectra of some rare-gas atoms ionized by the SXR alone since accurate photoionization cross sections of rare-gas atoms are available from experiments or theoretical calculations. To get information on the spectral phase $\Phi(\Omega)$, photoelectron spectra are generated by the same SXR in the presence of a time-delayed driving laser that is used to generate harmonics. The resulting electron spectra vs the time delay is called the spectrogram (or the streaking trace) and is then analyzed to retrieve the spectral phase. In this method, it is assumed that the spectrogram can be calculated using the so-called strong-field approximation (SFA)\cite{Mairesse_2005,Gagnon_apb2008}:

 \begin{eqnarray}
S(E,\tau)&=&\left|\int_{-\infty}^{\infty}E_{\text{SXR}}(t-\tau)d(p+A(t))\right.\nonumber\\
&&\left.\times{e^{-i\phi(p,t)}}e^{i\left(\frac{p^2}{2}+I_P\right)t}dt\right|^2,\label{eq_SFA}
\end{eqnarray}
where $E=p^2/2$ is the photoelectron energy, $\tau$ is the time delay between the two pulses, $A\left(t \right)$ is the vector potential of the laser, $d$ is the dipole transition matrix element, $I_p$ is the ionization potential of the atom, and $\phi(p,t)=\int_t^{\infty} [pA(t')+A^2(t')/2]dt'$ is the action of the electron in the laser field.

Atomic units are used throughout the paper unless otherwise stated. The vector potential $A$ should not be too large so it cannot contribute to the ionization of the target atom. In this model, the electron is removed by the SXR and then streaked by the vector potential of the laser until the pulse is over. To obtain the phase of $E_{\text{SXR}}$, previously the so-called FROG-CRAB~\cite{Mairesse_2005,Gagnon_apb2008} was used. It is based on the FROG method for retrieving the spectral phase of a femtosecond laser. To use the algorithm written for FROG, an additional approximation, called central momentum approximation (CMA), has to be made by replacing the $p$ in the action $\phi$ above by the momentum of the electron at the center of $U(\Omega)$. This approximation is not severe if the bandwidth is narrow, say about 10 to 20 eV. Thus, earlier attosecond pulses were retrieved using the FROG-CRAB method~\cite{Mairesse_2005,ZHC_2012OL}. It is generally believed that the method works for these narrowband  pulses.

To retrieve broadband pulses, the CMA has to be removed. Three methods have been proposed, PROOF~\cite{proof_2010}, VTGPA~\cite{VTGPA}, and PROBP~\cite{Zhao17_PRA}. The PROOF is based on approximating the action $\phi$ in Eq.~(1) by taking the limit when $A$ is small, and that the streaking IR field is a monochromatic wave. In PROOF, the streaking shift should be small, and thus it is more sensitive to errors in the streaking spectra. PROOF was used in~\cite{chang_natcom2017} and~\cite{ZHC_2012OL} even when the IR is a short few-cycle pulse. In the experiment of~\cite{Thomas_OE2017}, the spectral phase was retrieved using the ML-VTGPA method, which is a modification of VTGPA~\cite{VTGPA} to account for photoelectrons generated from multiple shells of the atom. In~\cite{Bieger_2017PRX}, the bandwidth of the SXR is about 200 eV. Using FROG-CRAB, the pulse duration obtained was about 24 as. Based on the attosecond lighthouse model, it was estimated that the pulse duration should be less than 322 as.

Among the phase retrieval algorithms, including FROG-CRAB, all  are based on iterative methods. The calculation is terminated after tens of thousands of iterations when the merit is not changing. To ``prove" that the retrieved SXR pulse (also for XUV pulse) is correct, the retrieved pulse is then used in Eq.~(1) to calculate the spectrogram. By comparing the experimental spectrogram with the one from the retrieved pulse visually, it is often deemed that the agreement is good. This procedure is probably acceptable for narrowband pulses, but not for broadband pulses.

\begin{figure*}[htbp]
\includegraphics[width=1.6\columnwidth,height=240pt,clip=true]{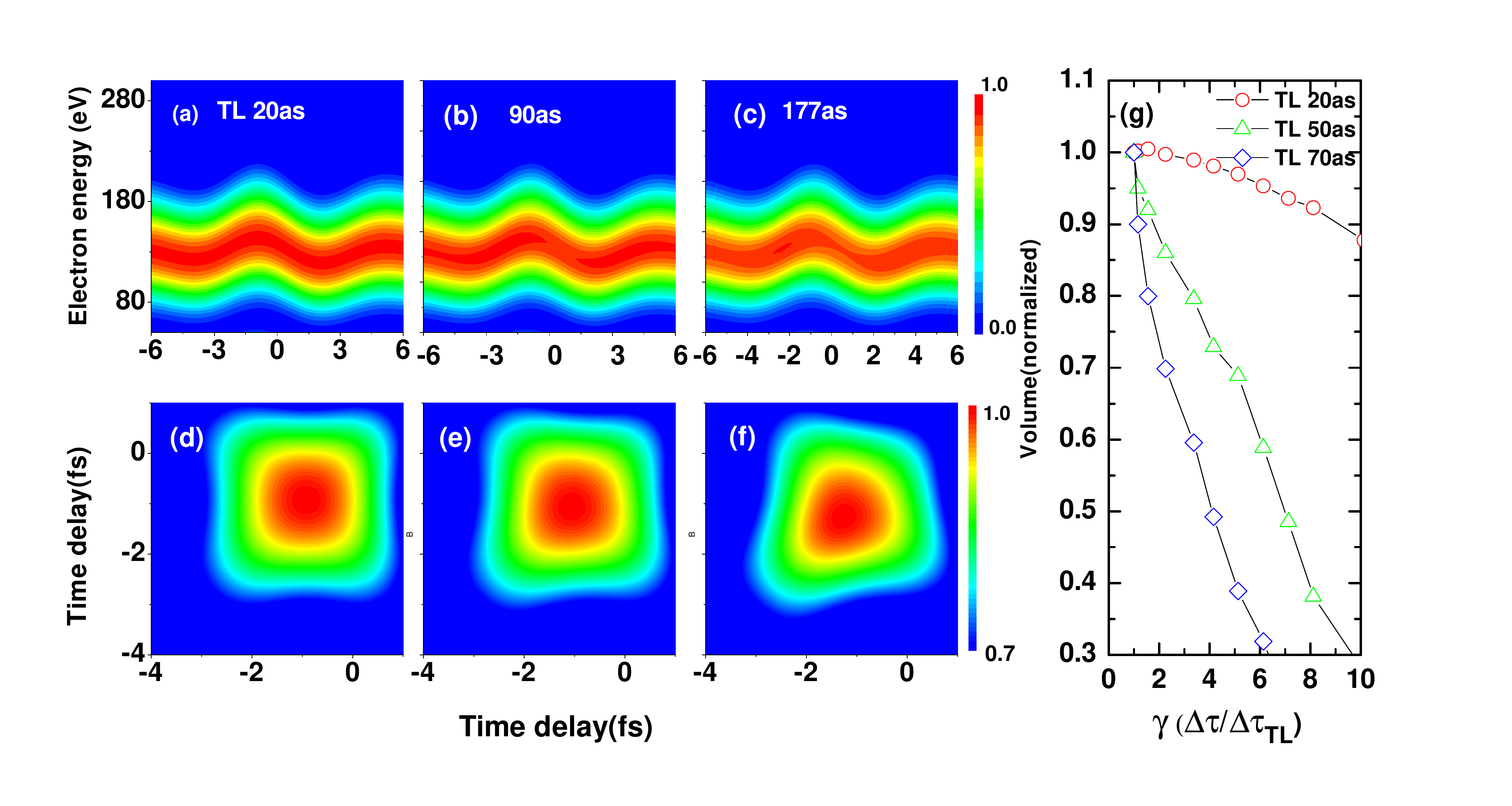}
\caption{(Color online) Theoretically calculated spectrograms (a,b,c) and the corresponding AC patterns (d,e,f). The central energy of the three XUV pulses is 160 eV, with a FWHM bandwidth of 94 eV. The duration of the TL pulse is 20 as, while the other two are 90 and 177 as, respectively. The MIR used in the simulation is 1800 nm in wavelength, 5.7 fs in FWHM duration and $2.5\times10^{12}$ W/cm$^{2}$ in peak intensity. (g) Normalized volume of the AC pattern vs  pulse duration in units of the duration of the transform-limited pulse. The normalized volume drops more slowly versus the scaled duration for broadband pulses than for narrowband pulses, making it harder to retrieve broadband pulses. The examples assume that the XUV pules have quadratic phases only; see text.\label{f1}}
\end{figure*}

To illustrate this point, we first compare how the spectrogram calculated using Eq.~(1) depends on the spectral phase. For simplicity, consider a SXR pulse in the energy domain, $E_{SXR}(\Omega)=U(\Omega)e^{i\Phi(\Omega)}$, with a Gaussian spectral amplitude $U(\Omega)=U_0 e^{-2\ln2\frac{(\Omega-\Omega_0)^2}{(\Delta\Omega)^2}}$ and a quadratic spectral phase $\Phi(\Omega)=a_{2}\frac{(\Omega-\Omega_{0})^{2}}{(\Delta\Omega/2)^{2}}$. Here we use $\Omega_{0}=164$ eV as the central photon energy and $\Delta\Omega=94$~eV as the full width at half maximum (FWHM) bandwidth, which can support a transform-limited (TL) pulse (corresponding to $a_2=0$) of 20~as. The coefficient $a_2$ is a measure of the attochirp, or equivalently we can use the parameter $\gamma=\Delta \tau/\Delta\tau_{\text{TL}}$, which is the ratio of the duration of the chirped pulse to the TL duration. Figures~1(a)-(c) compare the spectrograms simulated according to Eq.~(\ref{eq_SFA}). The mid-IR (MIR) used is 1800~nm in wavelength, 5.7~fs in duration, and $2.5\times10^{12}$ W/cm$^{2}$ in peak intensity. From Figs. 1(a)-(c), visually the three spectrograms show little difference, even though their pulse durations are 20, 90, and 177~as, respectively. To magnify their contrast, we calculate the auto-correlation (AC or $Q$) of the spectrogram, defined by

\begin{equation}\label{eq_Q}
Q(\tau_1,\tau_2)=\int_0^\infty S(E,\tau_1)S(E,\tau_2)dE.
\end{equation}
Figures 1(d)-(f) show the corresponding AC patterns over one optical period. For the TL pulse, the AC shape is close to a square in the center. As the linear chirp is increased, the square gradually deforms and skews along the diagonal axis. Clearly, the more the deformation is, the larger is the linear chirp (and pulse duration). To delineate the relative deformation of the AC with respect to the TL pulse, we define a normalized volume $V_\text{norm}$  for each AC pattern:

\begin{equation}\label{eq_v}
V_{norm}=\frac{\int \int Q({\tau_{1},\tau_{2}})d\tau_{1}d\tau_{2}}{\int \int Q^{TL}({\tau_{1},\tau_{2}})d\tau_{1}d\tau_{2}},
\end{equation}
in which we integrate the AC pattern (transform-limited AC pattern) over an area of half an optical cycle $T$ along each axis centered at the coordinate with the maximum value of the AC pattern (transformed-limited AC pattern). Figure~1(g) shows that V$_{\text{norm}}$ drops very quickly with $\gamma$ if the TL pulse is 70~as, but very slowly if the TL pulse is 20~as. This speaks that attosecond pulses with a broader bandwidth are much more difficult to retrieve. The MIR used for Fig.~1(g) has the wavelength of 1800 nm.


\begin{figure*}[htbp]
\includegraphics[width=1.8\columnwidth,height=320pt,clip=true]{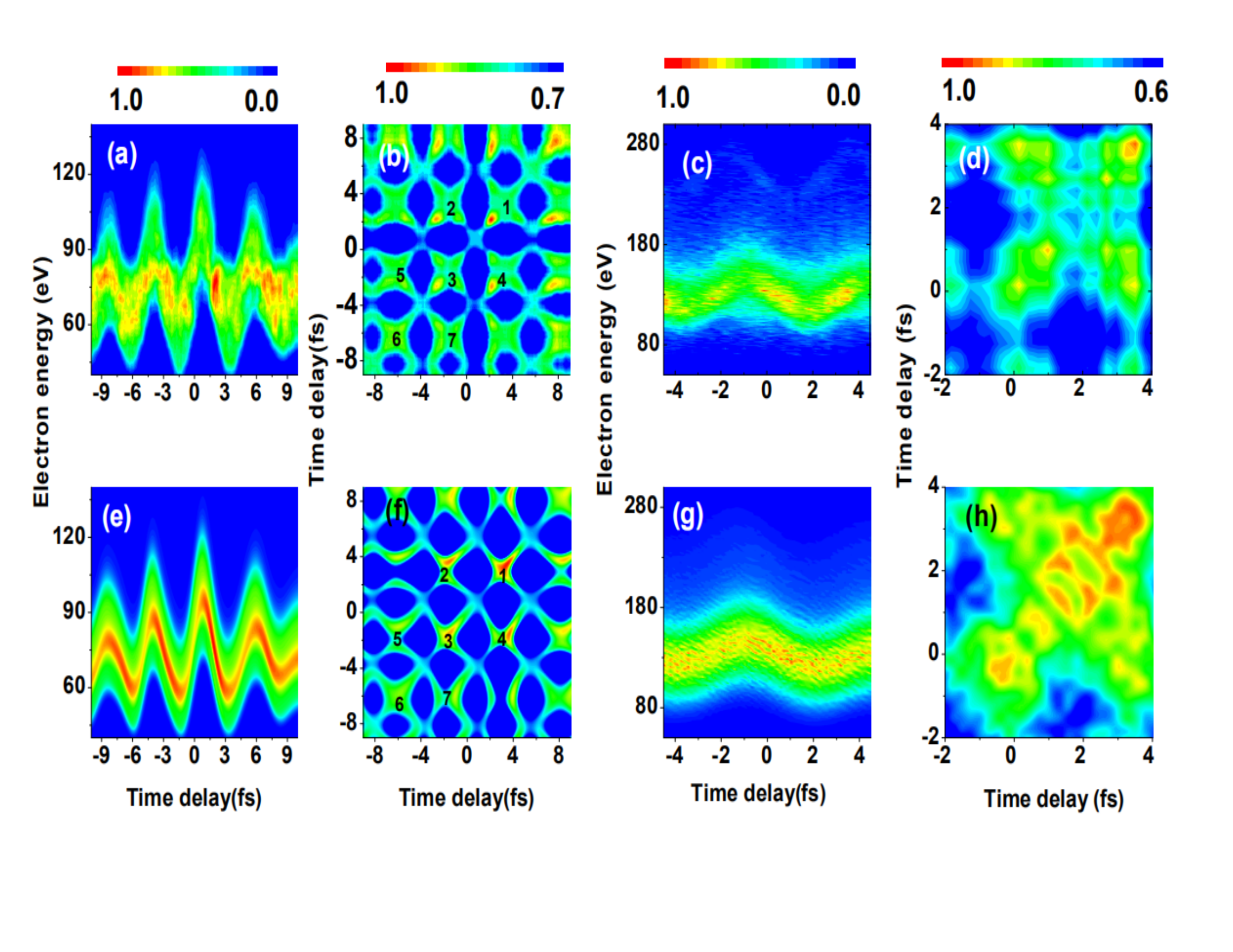}
\caption{  Experimental spectrograms and corresponding AC patterns. The two left columns are from ~\cite{Thomas_OE2017} and the two right columns are from \cite{chang_natcom2017}.
The upper row data are from the experiments and the lower row data are from the reconstructed theoretical data using the retrieved SXR and MIR pulses. Each pair along the column should match well if the retrieval is accurate.
\label{f3}}
\end{figure*}

To elaborate the advantage of using the AC instead of the spectrogram for spectral phase retrieval, in Figs. 2(a,e), we show the spectrograms from experiment with the one obtained from the retrieved pulse reported in ~\cite{Thomas_OE2017}. We calculate the ACs from these two spectrograms, and the results are shown in Figs. 2(b,f). The AC patterns are expected to repeat reasonably well for  each optical cycle of the MIR laser. We label several blocks (blocks 1-7), and one can see that the ACs from the experiment and from the simulated pulse do not agree well. Similar, in Figs. 2(c,g), the spectrograms from the experiment and from the simulated pulse of ~\cite{chang_natcom2017} are compared, and their corresponding ACs are compared in Figs. 2(d,h). Clearly, the latter shows that the experimental one and the retrieved one do not match well. Such discrepancy points that the SXR pulses reported in these experiments were not accurately retrieved. These results also demonstrate that the ACs serve as a good metric for evaluating the quality of the broadband SXR pulses retrieved, independent of whatever retrieval method is used.

The method we used to retrieve the spectral phase of the SXR pulse is called PROBP-AC~\cite{Yu2019_PRA}. It is a revision from the earlier proposed PROBP method~\cite{Zhao17_PRA}. Here, we first report the re-analysis of the data from ~\cite{Thomas_OE2017}, where the ML-VTGPA method was used.
From Fig.~2(a), the  spectrogram from ~\cite{Thomas_OE2017} appears to be well-behaved from $\tau=$ -9~fs to -1~fs. Thus we choose to analyze the AC in block number 5.[For more details, see sections I and III of the Supplementary Information (SI)]. The ACs for block number 5 from the experimental and the retrieved pulses from ~\cite{Thomas_OE2017} are shown in Figs. 3(a) and (b), respectively. We can see that they show little resemblance. Using our PROBP-AC method, the resulting AC is shown in Fig. 3(c). It is much closer to the experimental data given in Fig. 3(a). In our PROBP-AC method, the spectral amplitude of the SXR is obtained from the experiment. The vector potential of the 1800-nm MIR laser is also retrieved. From our results, we reconstruct the intensity of the SXR in the time domain, see Fig. 3(d). Our result does not agree very well with the one reported in ~\cite{Thomas_OE2017}. The FWHM pulse duration from our new evaluation is 62 as, as compared to 43 as reported in ~\cite{Thomas_OE2017}. In Fig. 3(e), we compare the spectral phases. The phase obtained in ~\cite{Thomas_OE2017} is very small over the whole spectral range, thus they retrieved a near-TL pulse. Our result from PROBP-AC shows large chirps away from the central energy. Note that the linear term in the spectral phase has been removed. Thus, the spectral phases presented in Fig. 3(e) show the phases that contribute to the pulse duration beyond the TL pulse. The reason for the discrepancy between the present result from ~\cite{Thomas_OE2017} is further discussed in Sec.II and III of SI. On the other hand, the better agreement of Fig. 3(c) than Fig. 3(b) with the experimental data Fig. 3(a) is a good indication that the present retrieved result is more accurate. We can also calculate the merit of the retrieved pulses (see Sec. IV of SI) using the AC or the spectrogram. From the AC, our (~\cite{Thomas_OE2017}) merit is 0.03 (0.09), and from the spectrogram is 0.010 (0.019), both showing our method obtains better merits. Fig.~3(f) compares the retrieved vector potential of the two methods. On femtosecond timescales the agreement between the two retrieved vector potentials is quite good.


\begin{figure*}[htbp]
\includegraphics[width=1.6\columnwidth,height=240pt,clip=true]{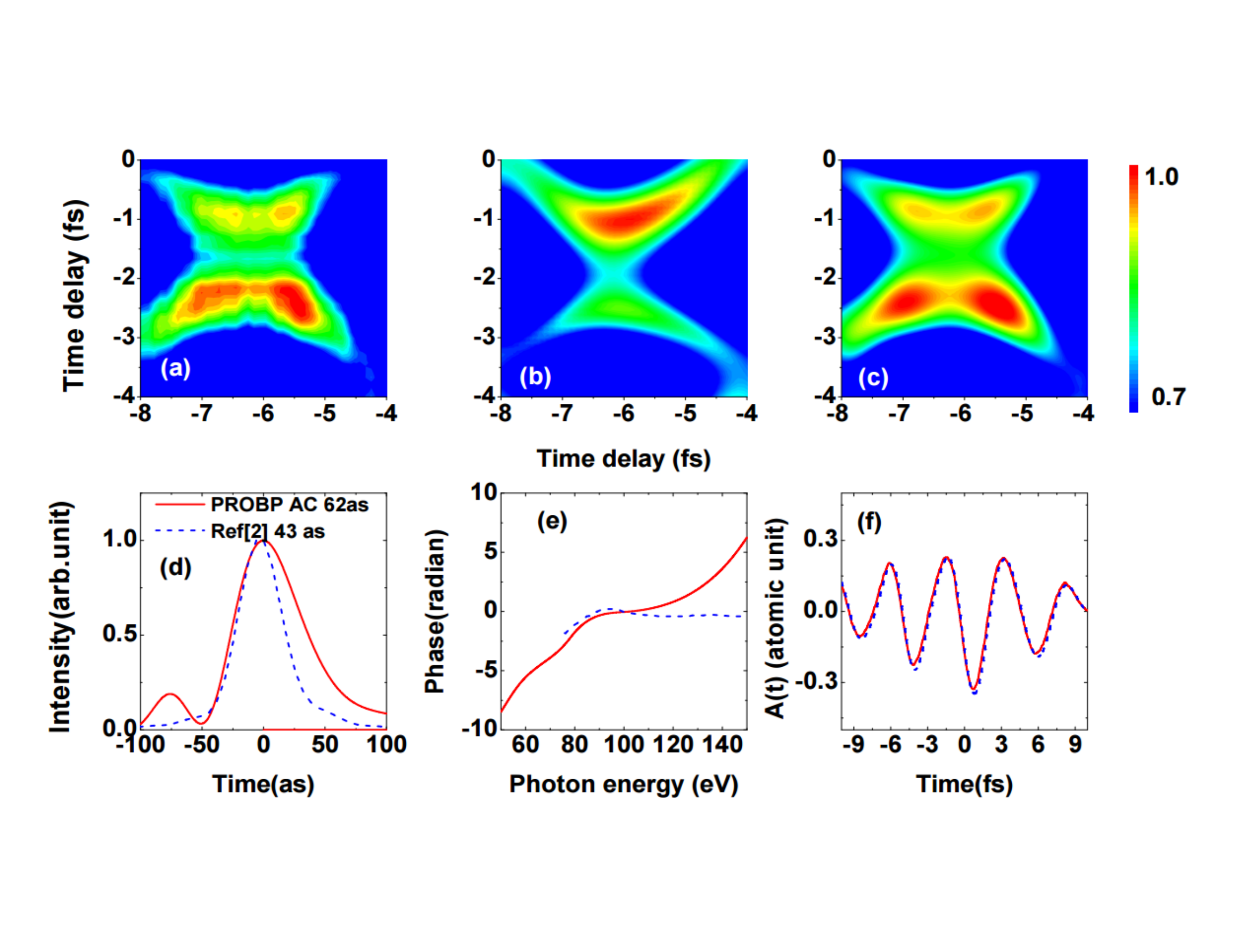}
\caption{(Color online) (a)Experimental AC pattern and, (b) the retrieved AC pattern from ~\cite{Thomas_OE2017}. (c) The retrieved AC pattern from the present method. (d) The retrieved temporal intensity envelope, (e) the spectral phase of the SXR, and (f) the vector potential of the MIR. The retrieval is obtained from the AC pattern of block 5 in Fig. 2(c). In (d-f), the solid red lines are from the present retrieval method and the blue dashed lines are from ~\cite{Thomas_OE2017}.
\label{f4}}
\end{figure*}

\begin{figure*}[htbp]
\includegraphics[width=1.6\columnwidth,height=240pt,clip=true]{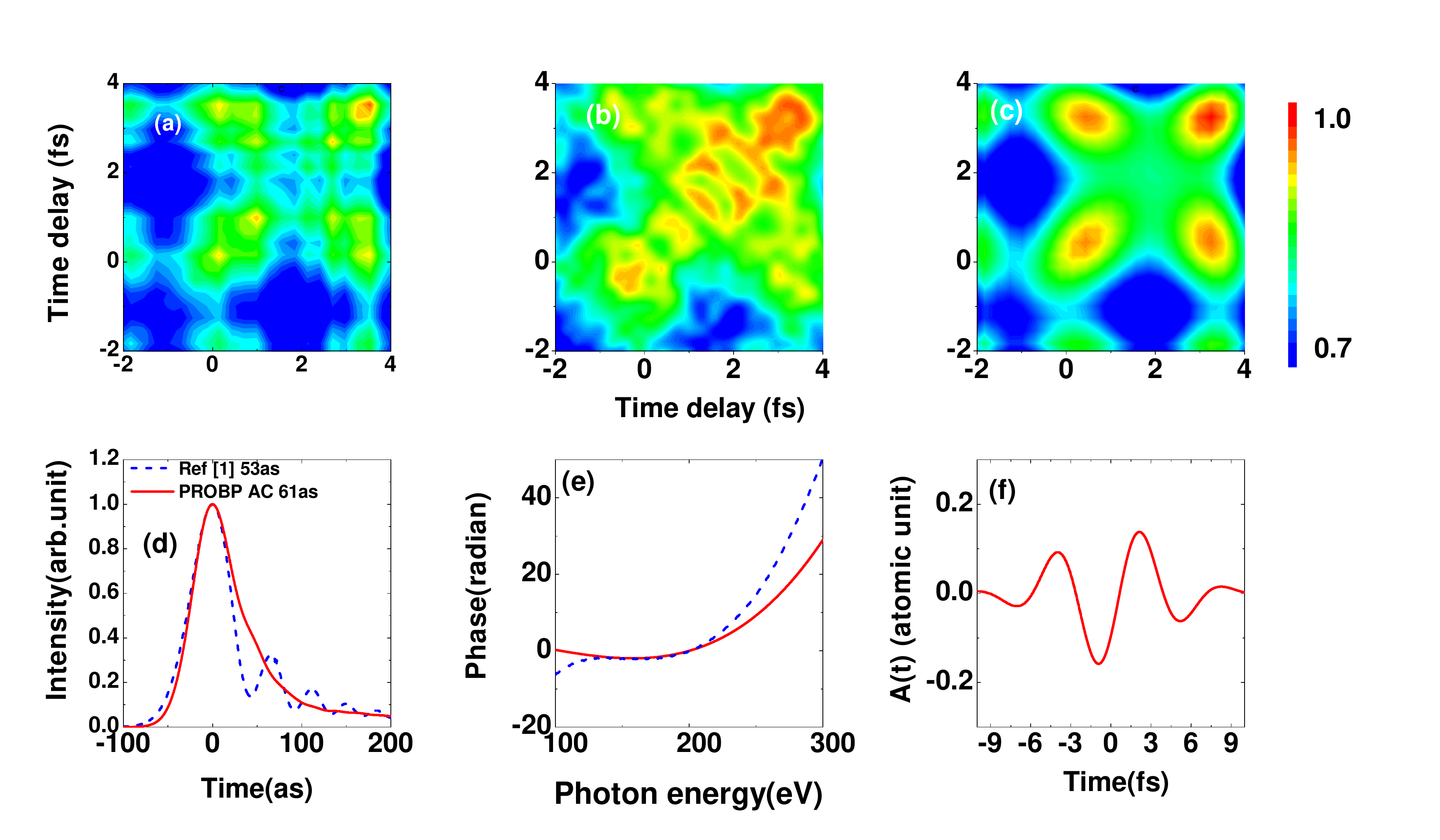}
\caption{(Color online) (a) and (b). AC patterns from the experimental and the retrieved spectrograms reported in Ref.~\cite{chang_natcom2017}, respectively. (c) Reconstructed AC pattern from the present method. (d)Temporal intensity envelopes of the SXR. (e) Spectral phase of the SXR retrieved and (f) vector potential of the MIR retrieved from the present method. In (d) and (e), the red solid lines are from the present method, and the blue-dashed lines are from the PROOF results of Ref.~\cite{chang_natcom2017}.
\label{f5}}
\end{figure*}

Next we re-examine the SXR pulse generated by a two-cycle pulse near 1.8 $\mu$m  reported in~\cite{chang_natcom2017}.  Their spectrum extends from about 100 to 300 eV;   the bandwidth is 94 eV and the TL pulse is 20 as. This pulse was retrieved in~\cite{chang_natcom2017} based on the PROOF method which approximates the SFA, Eq. (1),  under the condition that the streaking MIR is monochromatic and that the vector potential of the MIR is very weak. In this experiment, the target atom for the streaking spectra is helium. Fig.~4 summarizes the retrieved results of~\cite{chang_natcom2017} using PROOF and from our PROBP-AC method. Fig. 4(c) shows the AC from our retrieved result. It agrees better with the AC from the experimental data in \cite{chang_natcom2017} shown in Fig. 4(a) than the AC obtained from the retrieved SXR using PROOF (Fig. 4(b)). The pulse duration retrieved from our method is 61 as, while from PROOF it is 53$\pm 6$ as. Even though the pulse durations are quite close, the PROOF result shows a larger spectral phase change from 50 eV to 300 eV. Such large phase variation results in a structured long tail of the SXR pulse in the time domain, as seen in the blue-dashed line of Fig.~4(d). Note that the intensity of the second peak there is about 0.34 of the central peak, and thus the electric field at the second peak is 60$\%$ of the main peak. The result from PROBP-AC gives a smaller spectral phase, and a much better behaved SXR pulse in the time domain, see the red line in Fig. 4(d). The retrieved two-cycle MIR pulse is shown in Fig. 4(f) (The PROOF method assumes that the MIR is monochromatic). Based on the comparison of the ACs in Fig. 4, it is fair to say that the PROBP-AC result is more accurate. This is also supported by the merits between this work and the PROOF method, $0.036$ vs $0.061$ based on the ACs, or $0.012$ vs $0.032$ using the spectrograms (see sec IV of SI).

We also retrieved the streaking data from~\cite{Bieger_2017PRX}. The central photon energy of the SXR is at 250 eV, with a bandwidth of about 200 eV, corresponding to a TL pulse of about 10 as. Ionization of the SXR on Kr atoms would generate most of the photoelectrons from the $3d$. Fig. 5(a) shows the experimental spectrogram. Due to the weak signal, the spectrogram  does not show clean oscillation with respect to the optical period of the Mid-IR pulse. Since the spectrogram shows good oscillatory behavior from 15 to 35 fs in Fig. 5(a), we calculate the AC for this range, and the result is shown in Fig. 5(b). Using this AC pattern, we retrieve the spectral phase. The retrieved time-domain intensity profile is shown in Fig. 5(c). It has a FWHM pulse duration of 165 as, about half of the upper limit of 322 as reported in ~\cite{Bieger_2017PRX} based on the attosecond lighthouse model. The spectral phase obtained from our retrieval method is given in the inset of Fig. 5(c). Comparing the AC using the retrieved pulses, as shown in Fig. 5(d), it does show an overall global agreement with Fig.~5(b), but clearly the details are different. Due to the broader bandwidth and higher photon energies, the weaker signals and longer time (about 10 hrs) in collecting the spectrogram, it is not clear that the 165-as duration retrieved is the ``shortest" pulse that can be obtained for this broadband pulse. Since broader bandwidth would incur larger chirps in the spectral phase away from the center, a shorter TL pulse may not necessarily be a better way for generating a shorter attosecond pulse below 50 or 60 as.

\section{Discussion}
 One of the grand goals of ultrafast and attosecond physics is to generate even shorter light pulses in the soft X-ray region for probing inner-shell electron dynamics of materials. Often it was assumed that one would just have to keep generating continuum harmonics over ever increasing bandwidth. This would work if the spectral phase can be compensated over the whole broad energy region. As addressed in ~\cite{Bieger_2017PRX}, there is still no practical method available to do that.   In the meanwhile, accurate phase retrieval of a broadband pulse using the spectrogram has been shown to be very slowly converging. In this work, we demonstrated that phase retrieved directly from the autocorrelation (AC) of the spectrogram is more efficient and more accurate. We also demonstrated that correct phase is retrieved when the AC from the experimental data agrees with the AC from the retrieved pulse. The PROBP-AC method is expected to provide  the metrology of broadband pulses in the future.

\section{Methods}
The phase retrieval of broadband pulses (PROBP) was first introduced in \cite{Zhao17_PRA}, where the unknown laser and/or SXR are to be retrieved from the spectrogram. it does not impose the central momentum approximation. In PROBP, the spectral amplitude $U(\Omega)$ of the SXR is known from the experiment. The vector potential of the streaking MIR field is also expressed in the energy domain

\begin{equation}\label{eq_IRt}
A\left( \Omega  \right) = f\left( \Omega  \right){e^{  i\Psi \left( \Omega  \right)}}.
\end{equation}

Each of the unknown functions $\Phi(\Omega)$, $f\left( \Omega  \right)$, and $\Psi \left( \Omega  \right)$, respectively, is expanded in terms of B-spline basis functions

\begin{equation}\label{eq_fx}
f(x)=\sum_{i=1}^{n}g_i B_{i}^{k}(x).
\end{equation}

With some guessed parameters of these unknown functions, the constructed SXR and MIR are used in Eq.~(1) to obtain the spectrogram. By comparing the resulting spectrogram with the experiment, a genetic algorithm was used to select the new guesses for the next iteration. The iterative process is terminated after tens of thousands steps or after some preselected merit is reached. The PROBP method has been shown to work well for pulses with bandwidth up to about 100~eV~\cite{Zhao17_PRA}. The convergence becomes much slower for pulses with larger chirps or broader bandwidths. Since the AC appears to be a more sensitive marker of the spectral phase than the spectrogram, in the PROBP-AC method~\cite{Yu2019_PRA}, we retrieve the phase directly from the experimental AC patterns.

In the iterative method, we apply the genetic algorithm.  The fitness function is defined as the sum of
\begin{equation}\label{eq_error}
E = \sum\limits_{i,j} {\min \left( {{Q_{0}}(i,j),{Q_{1}}(i,j)} \right)} ,
\end{equation}
where ${Q_0}$ and ${Q_1}$ are the normalized input AC from the experiment and the reconstructed
AC, respectively. The $\min \left( {x,y} \right)$ is defined as the smaller  of $x$ and $y$. If the input   and reconstructed ACs are exactly the same, the fitness function is equal to $1$. In the numerical computation we discretize the
spectrogram $S(E,\tau)$ and the AC pattern $Q(\tau_1,\tau_2)$ on grid points. We use the genetic algorithm (GA) to find the optimal parameters  that would minimize Eq.~(\ref{eq_error}).

For narrower bandwidth pulses, previous studies~\cite{Yu2019_PRA} show that the PROBP-AC method converges much faster and more accurately. For the broadband pulses discussed here, we use the PROBP-AC method only since the PROBP method does not converge or take a long time to reach convergence. The amplitude and phase of the vector potential $A$ of the MIR field are also obtained. The experimental spectrogram is first normalized within the same time delay domain.

\begin{figure*}[!htbp]
\includegraphics[width=1.2\columnwidth,height=240pt,clip=true]{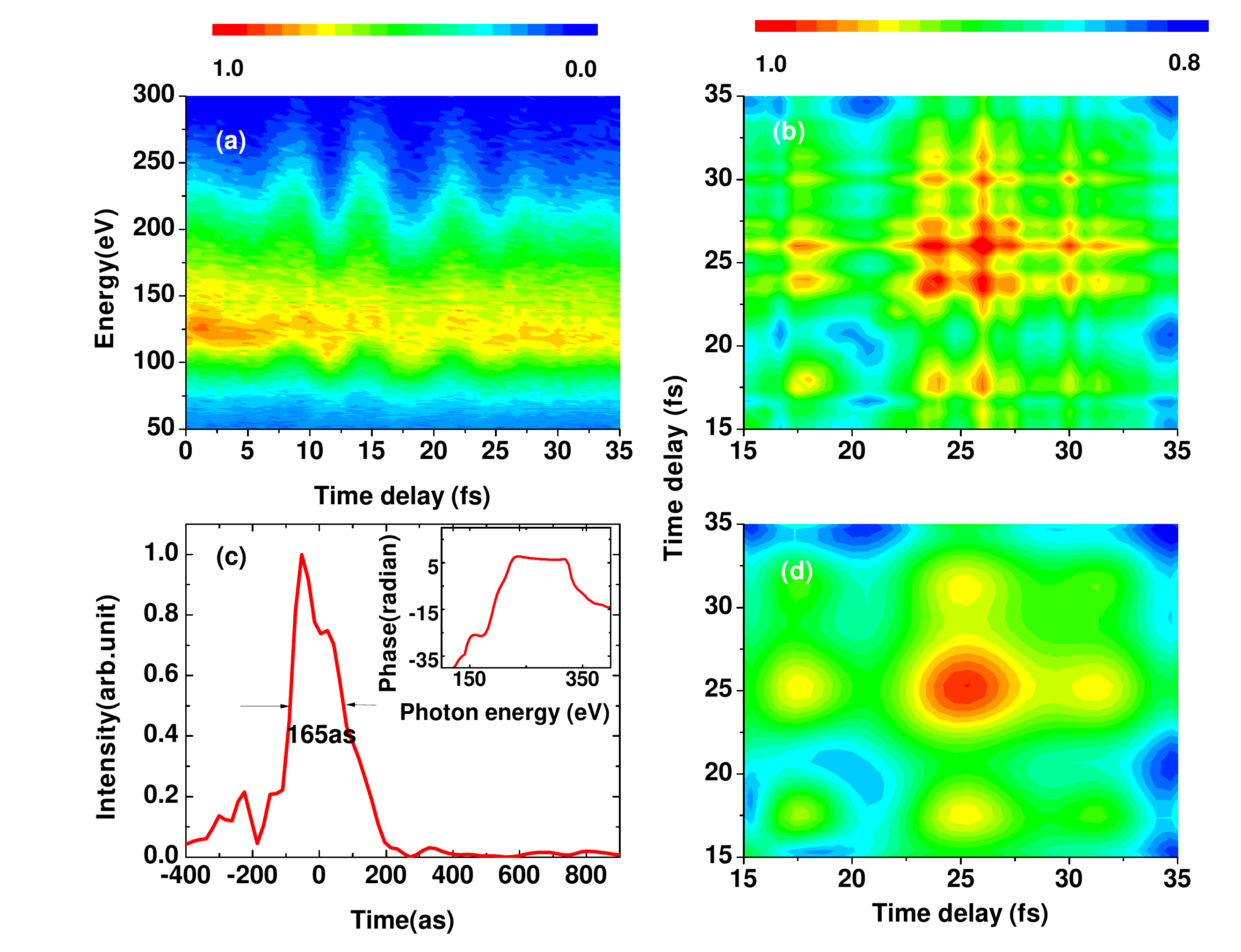}
\caption{(Color online)  (a) and (b). Experimental trace and the derived AC pattern  from~\cite{Bieger_2017PRX}, respectively. (d) The AC obtained from the present retrieved pulse. It is only in marginal agreement with the one in Fig. 5(b). (c) Envelope of the temporal intensity of the SXR; the inset is the spectral phase. Both are from the present retrieval method.
\label{f6}}
\end{figure*}

\section {Acknowledgement}
We thank Professors Hans Jakob Woerner, Zenghu Chang and Jens Biegert for providing the digital spectrograms from their experiments for us to analyze using our phase retrieval algorithm and for input to our initial draft of the manuscript. We also wish to thank Dr. P. D. Keathley for communicating some comparison with the result from the VTGPA method.
This research was supported in part by the Chemical
Sciences, Geosciences, and Biosciences Division, Office of
Basic Energy Sciences, Office of Science, US Department of
Energy, under Grant No. DE-FG02-86ER13491. W.Y. would
also like to acknowledge partial support by the Chinese Scholarship
Council (CSC), and by the National Natural Science
Foundation of China under Grant No. 11604131.


\begin{thebibliography}{0}
\expandafter\ifx\csname natexlab\endcsname\relax\def\natexlab#1{#1}\fi
\expandafter\ifx\csname bibnamefont\endcsname\relax
  \def\bibnamefont#1{#1}\fi
\expandafter\ifx\csname bibfnamefont\endcsname\relax
  \def\bibfnamefont#1{#1}\fi
\expandafter\ifx\csname citenamefont\endcsname\relax
  \def\citenamefont#1{#1}\fi
\expandafter\ifx\csname url\endcsname\relax
  \def\url#1{\texttt{#1}}\fi
\expandafter\ifx\csname urlprefix\endcsname\relax\def\urlprefix{URL }\fi
\providecommand{\bibinfo}[2]{#2}
\providecommand{\eprint}[2][]{\url{#2}}

\end{thebibliography}


\begin{thebibliography}{xx}


\bibitem{chang_natcom2017} Li, J. \textit{et al.} 53-attosecond X-ray pulses reach the carbon K-edge. Nat.~Commun. \textbf{8} 186 (2017).

\bibitem{Thomas_OE2017}Gaumnitz,~T.,~Jain, A.,~Pertot, Y.,~Huppert, M.,~Jordan, I.,~Lamas F.~A. \& W\"orner, H.~J.~ Streaking of 43-attosecond soft-X-ray pulses generated by a passively CEP-stable mid-infrared driver. Opt.~Exp. \textbf{25}, 27506-27518 (2017).


\bibitem{Bieger_2017PRX}Cousin, S. L.~ \textit{et al.} Attosecond Streaking in the Water Window: A New Regime of Attosecond Pulse Characterization. Phys.~Rev.~X \textbf{7}, 041030 (2017).

\bibitem{Krausz_2001nature}Hentschel, M. \textit{et al.} Attosecond metrology. Nature(London) \textbf{414}, 509-513 (2001).

\bibitem{Linbook}Lin, C.~D.,~Le, A.~T.,~Jin, C. \& Wei, H.~\emph{ Attosecond and Strong-Field Physics Principles and Applications} (Cambrige University press, 2018).

\bibitem{Krausz_2002nature}Drescher,~M. \textit{et al.} Time-Resolved Atomic Inner-Shell Spectroscopy. Nature (London) \textbf{419}, 803-807 (2002).


\bibitem{Krausz_2013nature} Schiffrin,~A. \textit{et al.} Optical-Field-Induced Current in Dielectrics. Nature (London) \textbf{493}, 70-74 (2013).



\bibitem{Dimauro_2008nat_phys} Colosimo,~P. \textit{et al.} Scaling strong-field interactions towards the classical limit. Nat. Phys. \textbf{4}, 386-389 (2008).


\bibitem{Stagira_2007OL} Vozzi, C. \textit{et al.} Millijoule-level phase-stabilized few-optical-cycle infrared parametric source. Opt. Lett. \textbf{32}, 2957-2959 (2007).

\bibitem{Kapteyn_2012science}Popmintchev, T. \textit{et al.} Bright coherent ultrahigh harmonics in the keV X-ray regime
from mid-infrared femtosecond lasers. Science \textbf{336}, 1287-1291 (2012).

\bibitem{Midorikawa_2008PRL}Takahashi, E. J., Kanai, T., Ishikawa, K. L., Nabekawa, Y. \& Midorikawa, K. Coherent water window x-ray generation by phase-matched high-order harmonic generation in neutral media. Phys. Rev. Lett. \textbf{101}, 253901 (2008).

\bibitem{Itatani_2014nat_phys} Ishii, N., Kaneshima, K., Kitano, K., Kanai, T., Watanabe, S. \& Itatani, J. Carrier-envelope phase-dependent high harmonic generation in the water window using few-cycle infrared pulses. Nat. Commun. \textbf{5}, 3331 (2014).


\bibitem{Biegert_2014OL}Cousin, S. L., Silva, F., Teichmann, S., Hemmer, M., Buades, B. \& Biegert, J. High-flux table-top soft X-ray source driven by sub-2-cycle, CEP stable, 1.85-¦Ìm 1-kHz pulses for carbon K-edge spectroscopy. Opt. Lett. \textbf{39}, 5383-5386 (2014).



\bibitem{Schmidt_2010APL}Schmidt, B. E. \textit{et al.} Compression of 1.8 ¦Ìm laser pulses to sub-two optical cycles with bulk material. Appl. Phys. Lett. \textbf{96}, 121109 (2010).

\bibitem{Kartner_2016JPB} Stein, G. J. \textit{et al.} Water-window soft X-ray high-harmonic generation up to the nitrogen K-edge driven by a kHz, 2.1 ¦Ìm OPCPA source. Journal of Physics B: Atomic, Molecular and Optical Physics \textbf{49}, 155601 (2016).

\bibitem{Biegert_2015Nat_comm} Silva, F., Teichmann, S. M., Cousin, S. L., Hemmer, M. \& Biegert, J. Spatiotemporal isolation of attosecond soft X-ray pulses in the water window. Nat. Commun. \textbf{6}, 6611 (2015).

\bibitem{Ishii_2014Nat_comm} Ishii, N. \textit{et al.} Carrier-envelope phase-dependent high harmonic generation in
the water window using few-cycle infrared pulses. Nat. Commun \textbf{5}, 3331
(2014).

\bibitem{Keathley_2016JPB} Keathley, P. D. \textit{et al.} Water-window soft x-ray high-harmonic generation up to
the nitrogen K-edge driven by a kHz, 2.1 ¦Ìm OPCPA source. J. Phys. B: At.
Mol. Opt. Phys. \textbf{49}, 155601 (2016).

\bibitem{Biegert_2015Nat_comm2} Teichmann, S. M., Silva, F., Cousin, S. L., Hemmer, M. \& Biegert, J. 0.5-keV
Soft X-ray attosecond continua. Nat. Commun. \textbf{7}, 11493 (2015).

\bibitem{Li_2016APL} Li, J. \textit{et al.} Polarization gating of high harmonic generation in the water
window. Appl. Phys. Lett. \textbf{108}, 231102 (2016).

\bibitem{Mairesse_2005}Mairesse, Y. \& Qu\'er\'e, F. Frequency-resolved optical gating for complete reconstruction of attosecond bursts. Phys.~Rev.~A \textbf{71}, 011401 (2005).

\bibitem{Gagnon_apb2008}Gagnon, J., Goulielmakis, E. \& Yakovlev, V.~S. The accurate FROG characterization of attosecond pulses from streaking measurements. Appl.~Phys.~B \textbf{92}, 25-32 (2008).

\bibitem{ZHC_2012OL} Zhao, K., Zhang, Q., Chini, M., Wu, Y., Wang X. \& Chang Z. Tailoring a 67 attosecond pulse through advantageous phase-mismatch. Optics Letters \textbf{37}, 3891-3893 (2012).


\bibitem{proof_2010}Chini, M., Gilbertson, S., Khan, S.~D. \& Chang Z. Characterizing ultrabroadband attosecond lasers. Opt.~Express \textbf{18}, 13006-13016 (2010).

\bibitem{VTGPA}Keathley, P.~D., Bhardwaj, S., Moses, J., Laurent, G. \& K\"artner F.~X. Volkov transform generalized projection algorithm for attosecond pulse characterization. New J.~Phys. \textbf{18}, 073009 (2016).

\bibitem{Zhao17_PRA} Zhao, X., Wei, H., Wu, Y. \& Lin, C. D. Phase-retrieval algorithm for the characterization of broadband single attosecond pulses. Phys.~Rev.~A \textbf{95}, 043407 (2017).

\bibitem{Yu2019_PRA} Yu, W. W., Zhao, X., Wei, H., Wang, S. J. \& Lin, C. D. Method for spectral phase retrieval of single attosecond pulses utilizing the autocorrelation of photoelectron streaking spectra. Phys.~Rev.~A \textbf{99}, 033403 (2019).



\end{thebibliography}
\end{document}